\documentclass[aps, prl, twocolumn, notitlepage, groupedaddress, 10pt, floatfix]{revtex4-2}
\usepackage[pdftex,plainpages=false,colorlinks=true,citecolor=blue,linkcolor=blue,urlcolor=blue,filecolor=green,bookmarksopen=true]{hyperref}
\usepackage{amsmath}
\usepackage[caption = false]{subfig}
\usepackage{graphicx,epstopdf}
\usepackage[english]{babel}
\usepackage{blindtext}
\usepackage{lipsum}
\usepackage{amsfonts}
\usepackage{amssymb}
\usepackage{bbm}
\usepackage{enumerate}
\usepackage{color}
\usepackage{latexsym}
\usepackage{times,txfonts}
\usepackage{physics}

\DeclareSymbolFont{CMletters}{OML}{cmm}{m}{it}
\DeclareMathSymbol{J}{\mathalpha}{CMletters}{`J}
\DeclareMathSymbol{j}{\mathalpha}{CMletters}{`j}
\DeclareMathSymbol{U}{\mathalpha}{CMletters}{`U}

\begin{document}

\title{Dark-State-Induced Heat Rectification}

\author{Kasper Poulsen}
\email{poulsen@phys.au.dk}
\affiliation{Department of Physics and Astronomy, Aarhus University, Ny munkegade 120, 8000 Aarhus C, Denmark}

\author{Nikolaj T. Zinner}
\email{zinner@phys.au.dk}
\affiliation{Department of Physics and Astronomy, Aarhus University, Ny munkegade 120, 8000 Aarhus C, Denmark}

\begin{abstract}
\vspace{0.3cm}
\noindent Heat and noise control is essential for the continued development of quantum technologies. For this purpose, a particularly powerful tool is the heat rectifier, which allows for heat transport in one configuration of two baths but not the reverse. Here we propose a class of rectifiers that exploits the unidirectionality of a low temperature bath to force the system into a dark state thus blocking heat transport in one configuration of the two baths. However, if the two baths are switched around, a heat current is observed. An implementation using a qutrit coupled to two harmonic oscillators is proposed and rectification values beyond $10^3$ are achieved for realistic parameter values. Furthermore, we show that the heat current can be amplified by an order of magnitude through external driving without diminishing the diode functionality. The heat rectification effect is seen for a large range of parameters, and it is robust towards both decay and dephasing.
\vspace{0.4cm} 
\end{abstract}

\maketitle

\section{Introduction}

With the advance of the second quantum revolution and an ever increased ability to exploit quantum effects, it is pertinent to understand and control noise/heat flow in modern quantum devices. 
It has even been proposed that components like heat transistors \cite{PhysRevLett.116.200601, PhysRevB.101.184510, PhysRevE.98.022118} and heat diodes \cite{RevModPhys.84.1045, PhysRevB.98.035414, Tesser_2022} could be used for a heat based computer.
Unsurprisingly, this increased attention has resulted in a leap in understanding of both theoretical and experimental aspects of, e.g., heat engines \cite{PhysRevLett.123.240601, PhysRevLett.125.166802, Josefsson2018} and information engines \cite{PhysRevLett.113.030601, PhysRevLett.121.030604, poulsen2021non}.

A particularly useful device for controlling heat is the heat rectifier: a device that exhibits asymmetric transport of heat similar to the diode in electronics. Within the framework of boundary-driven quantum systems, a quantum system is coupled to two heat baths at the extremities \cite{landi2021non, PhysRevLett.106.217206, PhysRevLett.126.077203}. The diode properties of the quantum system can then be studied by calculating the heat transport for both configurations of the baths. Rectification has been found in a diverse set of models ranging from one \cite{PhysRevB.73.205415, Senior2020, PhysRevB.103.155434} or two \cite{PhysRevE.89.062109, PhysRevApplied.15.054050} two-level systems to large 1D spin chains \cite{PhysRevB.79.014207, PhysRevLett.120.200603, PhysRevB.80.172301, PhysRevE.99.032136, PhysRevE.102.062146} and 2D spin chain geometries \cite{PhysRevE.103.032108}. It has even been proposed to use quantum entanglement for enhanced rectification \cite{poulsen2021entanglement}. While large rectification factors have been found theoretically, recent proposals either require very large systems or are sensitive to decoherence \cite{poulsen2021entanglement,  PhysRevE.105.024120}. 

Going from a qubit to a qutrit offers many additional engineering opportunities such as dark states \cite{RevModPhys.77.633, PhysRevApplied.14.024092, PhysRevA.102.011502, PhysRevA.102.023707}. With a large anharmonicity, the baths can be engineered to only promote transitions between a specific pair of levels \cite{D_az_2021, PhysRevB.101.184510, poulsen2021non}. At the same time, qutrits remain simple to construct since many qubits have additional levels \cite{doi:10.1063/1.5089550, PhysRevResearch.2.032025, PhysRevE.100.032107, PhysRevB.99.144304}, and therefore, they offer a great platform for studying open system dynamics and quantum thermodynamics.





\section{Idealized version}
\begin{figure}[t]
\centering
\includegraphics[width=1 \linewidth, angle=0]{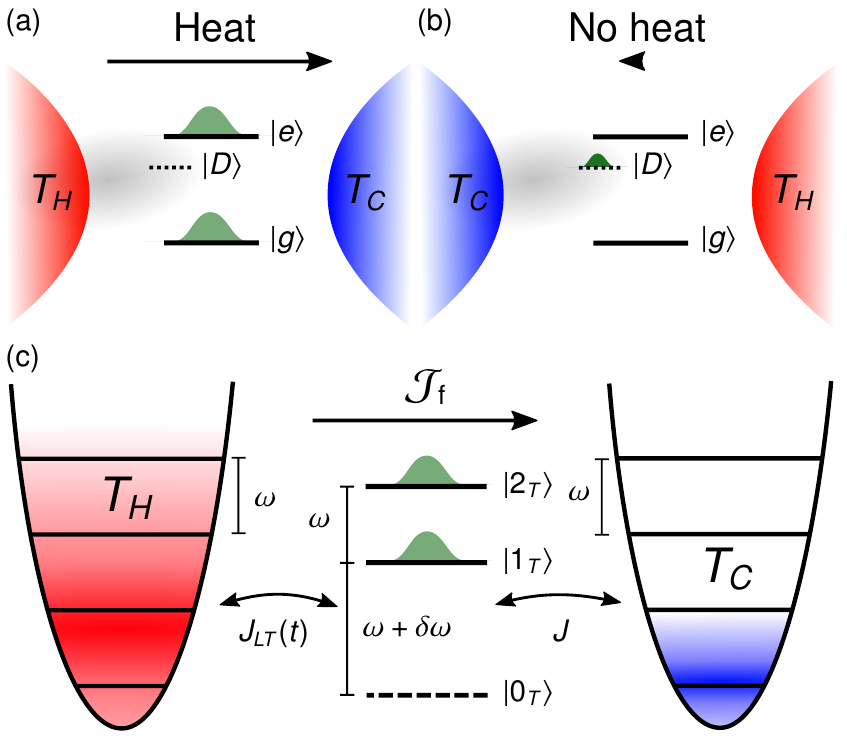}
\caption{(a)-(b) Minimal model for dark-state-induced rectification. (a) In forward bias, excitations are transported through the channel $|g\rangle \leftrightarrow |e\rangle$. (b) In reverse bias, the diode is closed by the dark state $|D\rangle$. (c) Possible model for dark-state-induced rectification consisting of a qutrit and two harmonic oscillators. The ground state acts as the dark state which is isolated from the right bath due to energy conservation.}
\label{figure1}
\end{figure}

Here we propose a class of rectifiers consisting of three states coupled to two baths as seen in Figs.~\ref{figure1}(a)-(b). In forward bias, the left bath is hot and the right bath is cold as seen in Fig.~\ref{figure1}(a), while reverse bias is seen in Fig.~\ref{figure1}(b). The system is engineered to exploit the unidirectionality of a cold bath to drive the system into the state $\ket{D}$ in reverse bias. $\ket{D}$ is a dark state of the right bath, and if it is populated, it will completely block any transport between the two baths. In forward bias, the two remaining states $\ket{g}$ and $\ket{e}$ facilitate transport as usual, thus implementing a perfect heat diode. The mechanism can easily be understood through the master equation
$\partial_t \vec{P}(t) = W \vec{P}(t)$ where $\vec{P} = [P(\ket{D}), P(\ket{g}), P(\ket{e})]^T$ is a vector of populations and $W$ is a matrix of rates which generally takes the form
\begin{align*}
W= \begin{pmatrix}
-\Gamma_{D\rightarrow g}-\Gamma_{D\rightarrow e} & \Gamma_{g \rightarrow D} & \Gamma_{e \rightarrow D}\\
\Gamma_{D\rightarrow g} & -\Gamma_{g \rightarrow D}- \Gamma_{g \rightarrow e} & \Gamma_{e \rightarrow g}\\
\Gamma_{D\rightarrow e} & \Gamma_{g \rightarrow e} & -\Gamma_{e \rightarrow D}-\Gamma_{e \rightarrow g}
\end{pmatrix} .
\end{align*}
Here $\Gamma_{a\rightarrow b}$ is the transition rate from $\ket{a}$ to $\ket{b}$. The left bath interaction is engineered such that, in reverse bias, the cold bath allows for transitions into the dark state but not out of it, i.e., $\Gamma_{D\rightarrow g}, \Gamma_{D\rightarrow e} =0$ and $\Gamma_{g\rightarrow D}, \Gamma_{e\rightarrow D} > 0$. The other rates are kept general for now, but an example is given in Eq.\eqref{rates}. In reverse bias, the rate matrix becomes
\begin{align*}
W= 
\begin{pmatrix}
0 & \Gamma_{g \rightarrow D} & \Gamma_{e \rightarrow D}\\
0 & -\Gamma_{g \rightarrow D}- \Gamma_{g \rightarrow e} & \Gamma_{e \rightarrow g}\\
0 & \Gamma_{g \rightarrow e} & -\Gamma_{e \rightarrow D}-\Gamma_{e \rightarrow g}
\end{pmatrix} .
\end{align*}
In the long time limit, the system will go towards the steady-state solution $\partial_t \vec{P}_\text{ss} = 0$, which is easily found to be
\begin{align*}
P_\text{ss} (\ket{D}) = 1.
\end{align*}
With one state fully populated no heat is transferred, and the system implements a perfect diode. To discuss the heat transferred in forward bias and to discuss one possible implementation of this idea, the example seen in Fig.~\ref{figure1}(c) is studied.

\section{Setup}

The Hamiltonian for the implementation shown in Fig.~\ref{figure1}(c) is
\begin{align*}
\hat{H} &= \omega \Big(\hat{a}_L^\dag \hat{a}_L + \hat{a}_R^\dag \hat{a}_R\Big) + (\omega +\delta \omega) \hat{a}_T^\dag \hat{a}_T  - \frac{\delta \omega}{2} \hat{a}_T^\dag \hat{a}_T \Big(\hat{a}_T^\dag \hat{a}_T -1\Big) \\ &\hspace{0.8cm} + J_{LT}(t) \Big(\hat{a}_L + \hat{a}_L^\dag \Big) \Big(\hat{a}_T + \hat{a}_T^\dag \Big) + J \Big(\hat{a}_T + \hat{a}_T^\dag \Big) \Big(\hat{a}_R + \hat{a}_R^\dag \Big) ,
\end{align*}
where the subscripts L, T, and R are used for the left harmonic oscillator, the qutrit, and the right harmonic oscillator, respectively. $\hat{a}_\alpha$ and $\hat{a}^\dag_\alpha$ for $\alpha \in \{L, T, R\}$ are ladder operators. $\omega$ is the frequency of the oscillators, and $\delta \omega$ is the anharmonicity of the qutrit. The left hopping is time dependent $J_{LT} (t) = J + J' \cos(\delta \omega t)$ and $J$ sets the energy scale of the system. We are using units where $\hbar = k_B=1$. Transforming into the interaction picture with respect to $\hat{H}_0 = \omega \sum_{\alpha \in \{L, T, R\}} \hat{a}_\alpha^\dag \hat{a}_\alpha$, performing a rotating wave approximation, and concatenating the qutrit to the three lowest levels  the Hamiltonian becomes
\begin{align*}
\hat{H}_I &= -\delta \omega \op{0_T} + J_{LT}(t) \Big(\hat{a}_L \hat{a}_T^\dag + \hat{a}_L^\dag \hat{a}_T \Big) + J \Big(\hat{a}_T \hat{a}_R^\dag + \hat{a}_T^\dag \hat{a}_R \Big), 
\end{align*}
where we have introduced the notation $\ket{0_T}$, $\ket{1_T}$, and $\ket{2_T}$ for the three qutrit states. The two harmonic oscillators act as the two bath in Figs.~\ref{figure1}(a)-(b), Due to the baths the system is open and its state is described through the density matrix $\hat{\rho}$. The decay of the harmonic oscillator correlation functions is modeled through the Lindblad master equation \cite{Lindblad1976, breuer2002theory}
\begin{align}
\frac{d \hat{\rho}}{d t} = \mathcal{L}[\hat{\rho}] &= -i [\hat{H}, \hat{\rho}] + \mathcal{D}_L[\hat{\rho}] + \mathcal{D}_R[\hat{\rho}], \label{me:1}
\end{align}
where $[\bullet, \bullet]$ is the commutator, $\mathcal{L}[\hat{\rho}]$ is the Lindblad superoperator, and $\mathcal{D}_{L(R)}[\hat{\rho}]$ is a dissipative term describing the action of the left (right) bath
\begin{align}
\mathcal{D}_{L(R)}[\hat{\rho}] &= \Gamma (n_{L(R)} + 1) \left( \hat{a}_{L(R)} \hat{\rho} \hat{a}_{L(R)}^\dag - \frac{1}{2} \{ \hat{a}_{L(R)}^\dag \hat{a}_{L(R)}, \rho \} \right) \\
& \hspace{1.5cm}+ \Gamma n_{L(R)} \left( \hat{a}_{L(R)}^\dag \hat{\rho} \hat{a}_{L(R)} - \frac{1}{2} \{ \hat{a}_{L(R)} \hat{a}_{L(R)}^\dag, \rho \} \right). \nonumber
\end{align}
where $\{\bullet, \bullet\}$ is the the anti-commutator. $\Gamma$ is the coupling strength between the baths and harmonic oscillators and $n_{L(R)}~=~\big(e^{\omega/T_{L(R)}}-1\big)^{-1}$ is the mean number of excitation in the left (right) harmonic oscillator in the absence of the qutrit. By forward bias, we denote the case where $n_L = n_H$ and $n_R = n_C$, and heat flows from left to right. By reverse bias, we denote the case where $n_L = n_C$ and $n_R = n_H$, and heat flows from right to left. We assume that $n_H > n_C$. After sufficient time the system will reach steady-state, $\mathcal{L}[\hat{\rho}_\text{ss}] = 0$. Unless otherwise stated, we use $\delta \omega = 300J$, $J'=0.5J$, $\Gamma = 10J$, $n_H = 0.5$, and $n_C = 0$. To study transport, we write the change in total energy of the system
\begin{align}
0=\frac{d\ev{H}_\text{ss}}{dt}  = \ev{\frac{d\hat{H}}{dt} }_\text{ss}  + \tr \left\{\hat{H} \mathcal{D}_L [\hat{\rho}_\text{ss}] \right\} + \tr \left\{\hat{H} \mathcal{D}_R [\hat{\rho}_\text{ss}] \right\}, \label{eq:heat}
\end{align}
where $\ev{\bullet}_\text{ss}= \tr \left\{\bullet \hat{\rho}_\text{ss} \right\}$ is the steady state expectation value, and $\tr \{\bullet\}$ is the trace over the entire Hilbert space. The first part is identified as the work, the second term is the heat increase due to the left bath, and the third term is the heat increase due to the right bath. From this, we define the two transport measures:
\begin{subequations}
\begin{alignat}{1}
\mathcal{J}_{L(R)} &= \Gamma n_{L(R)} \ev{\hat{a}_{L(R)} \hat{a}_{L(R)}^\dag}_\text{ss} - \Gamma (n_{L(R)}+1) \ev{\hat{a}_{L(R)}^\dag \hat{a}_{L(R)}}_\text{ss}, \label{eq:current} \\
\mathcal{W} &= \frac{J' \delta \omega}{2i} \ev{ \hat{a}_L \hat{a}_T^\dag e^{-i \delta \omega t} - \hat{a}_L^\dag \hat{a}_T e^{i \delta \omega t} }_\text{ss} .
\end{alignat}
\end{subequations}
Since the steady-state density matrix is independent of $\omega$, we use the $\omega$-independent excitation current $\mathcal{J}_{L(R)}$ instead of the heat current $\omega \mathcal{J}_{L(R)}$. Likewise, we will focus on the number of excitation added through work $\mathcal{W}/\delta \omega$. We have used that $\omega, \delta \omega \gg J, J'$ such that the substitution $\hat{H} \rightarrow \hat{H}_0$ can be made in the second and third term in Eq.~\eqref{eq:heat}. We define the forward bias excitation current to be $\mathcal{J}_\textrm{f} = -\mathcal{J}_R$, while the reverse bias excitation current is $\mathcal{J}_\textrm{r} = \mathcal{J}_L$. The quality of the diode is quantified using the rectification
\begin{align*}
\mathcal{R} = -\frac{\mathcal{J}_\textrm{f}}{\mathcal{J}_\textrm{r}},
\end{align*}
which tends to infinity for a perfect diode. 
In summary, the system is designed using the methodology seen in Figs.~\ref{figure1}(a)-(b). This is seen through the equivalences
\begin{align*}
\ket{D} \equiv \ket{0_T}, \quad \ket{g} \equiv \ket{1_T}, \quad \ket{e} \equiv \ket{2_T}.
\end{align*}
In forward bias, excitations can propagate through the system through, e.g., transitions like
\begin{align*}
\text{Forward:}& \quad \ket{1_L 1_T 0_R} \leftrightarrow \ket{0_L 2_T 0_R} \leftrightarrow \ket{0_L 1_T 1_R}.
\end{align*}
In reverse bias, the qutrit is trapped in the dark state $\ket{0_T}$ through the transitions
\begin{align*}
\text{Reverse:}& \quad \ket{0_L 1_T 0_R} \leftrightarrow \ket{1_L 0_T 0_R} \rightarrow \ket{0_L 0_T 0_R}.
\end{align*}
The first part is allowed when $J' > 0$, and the second part is due to the cold bath. When the qutrit is in the dark state, excitations are not allowed to propagate from the hot bath to the qutrit due to energy conservation.

\section{Results} 
\begin{figure}[t]
\centering
\includegraphics[width=1 \linewidth, angle=0]{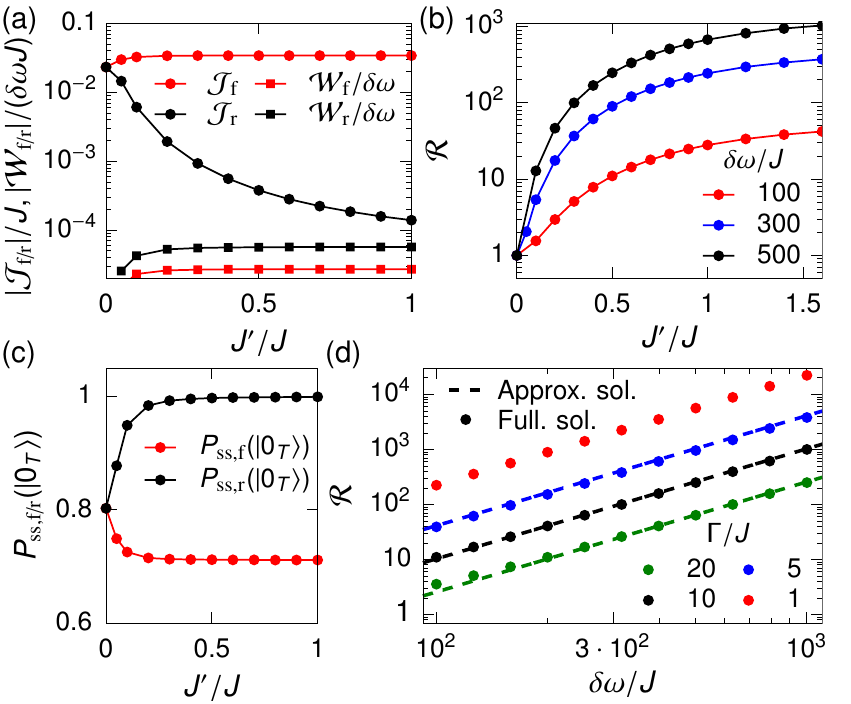}
\caption{(a) Excitation current and work as the diode is turned on through $J'$. (b) Rectification $\mathcal{R}$ as a function of $J'$ for different anharmonisities $\delta \omega$. (c) Steady-state population of the dark state as a function of $J'$ for both forward and reverse bias. (d) Rectification as a function of the anharmonisity $\delta \omega$ for different values of $\Gamma$. The full solution is plotted using points, and the approximate solution, Eq.~\eqref{eq:app_rec}, is plotted using a dashed line.}
\label{figure2}
\end{figure}

The excitation current and work results are plotted in Fig.~\ref{figure2}(a) as the diode is turned on via an increase in $J'$. The current, in reverse bias, becomes suppressed, and the number of excitations added through work per unit of time is orders of magnitude smaller than the excitation current in either bias. This is further verified by Fig.~\ref{figure2}(b) where the rectification is plotted. To verify that the current in reverse bias is indeed blocked due to the dark state being populated, we plot the dark state population, $P_{\text{ss}}(\ket{0_T}) = \ev{\op{0_T}}_{\text{ss}}$ in Fig.~\ref{figure2}(c). The dark state population in reverse bias does indeed approach unity as the diode is turned on.

\section{Markovian solution} 
An analytic solution can be found for $\omega, \delta \omega \gg \Gamma \gg J,J'$, $J'/J\gg \Gamma/\delta \omega$, $n_C=0$, and $n_H \leq 1$ \cite{poulsen2021non, PhysRevLett.128.240401}. In this regime, all coherences in $\hat{\rho}$ will decay rapidly, and the harmonic oscillators can be seen as baths with Lorenzian spectral densities through the Markov approximation which for the right bath becomes
\begin{align*}
S_{\hat{B}}(\omega') = \frac{(1+n_R) J^2 \Gamma }{(\omega - \omega')^2 + \Gamma^2/4} + \frac{n_R J^2 \Gamma }{(\omega + \omega')^2 + \Gamma^2/4}
\end{align*}
Here $\hat{B} = J\big(\hat{a}_R^\dag + \hat{a}_R \big)$ is the right harmonic oscillator operator that couples to the qutrit. The spectral density for the left harmonic oscillator is similar but with four terms due to the driving. Using this, the populations for the qutrit, $P(\ket{\alpha_T}) = \tr \{\op{\alpha_T} \hat{\rho}\}$, can be written in the form discussed earlier, $\partial_t \vec{P}(t) = W \vec{P}(t)$. The transition rates become
\begin{subequations}
\label{rates}
\begin{alignat}{1}
\Gamma_{2_T \rightarrow 0_T} = 0, \quad \Gamma_{1_T \rightarrow 0_T} = \frac{(1+n_L) J'^2 }{\Gamma} + \frac{(1+n_R ) J^2 \Gamma}{\delta \omega^2 + \Gamma^2/2}, \hspace{0.55cm}\\
\Gamma_{2_T \rightarrow 1_T} = \frac{8(1+n_L) J^2}{\Gamma} + \frac{8(1+n_R) J^2}{\Gamma},\quad \Gamma_{0_T \rightarrow 2_T} = 0, \hspace{0.5cm}\\
\Gamma_{0_T \rightarrow 1_T} = \frac{n_L J'^2 }{\Gamma} + \frac{n_R J^2 \Gamma}{\delta \omega^2 + \Gamma^2/2}, \,\, \Gamma_{1_T \rightarrow 2_T} = \frac{8n_L J^2}{\Gamma} + \frac{8n_R J^2}{\Gamma}. 
\end{alignat}
\end{subequations}
The first term in each rate is due to the left bath, while the second term is due to the right term. The solution to $W \vec{P}_{\text{ss}}=0$ can be found to be
\begin{equation}
\vec{P}_{\text{ss}} = \mathcal{N} \begin{pmatrix}
(2+n_R+n_L) \Big((1+n_L) J'^2 + (1+n_R) J^2 \frac{\Gamma^2}{\delta \omega^2}\Big) \\
(2+n_R+n_L) \Big( n_L J'^2 + n_R J^2 \frac{\Gamma^2}{\delta \omega^2}\Big) \\
(n_R + n_L)\Big(n_L J'^2 + n_R J^2 \frac{\Gamma^2}{\delta \omega^2}\Big)
\end{pmatrix}, \label{eq:steadystate_mark}
\end{equation}
where $\mathcal{N}$ is a constant ensuring $P_{\text{ss}}(\ket{0_T}) + P_{\text{ss}}(\ket{1_T}) + P_{\text{ss}}(\ket{2_T})=1$, and we have used the assumption $\delta \omega \gg \Gamma$. The current is now found as the number of excitations decaying due to the cold bath, e.g., $\mathcal{J}_{\textrm{f}} \simeq P(\ket{2_T}) \frac{8J^2}{\Gamma}$. Since one quanta of energy $\delta \omega$ is added through work every time the left bath causes a transition between $\ket{0_T}$ and $\ket{1_T}$, the excitation work is found as the number of excitations exchanged with the left bath through the $J'$-interaction, e.g., $\mathcal{W}_\textrm{r}/\delta \omega =- P(\ket{1_T}) \frac{J'^2}{\Gamma}$. Under the stated assumptions the current and work become
%
\begin{align*}
\mathcal{J}_\textrm{f} &= \frac{8 n_H^2}{2+5n_H+3n_H^2} \frac{J^2}{\Gamma},\\
\mathcal{J}_\textrm{r} &= -\frac{n_H}{2+n_H} \frac{8n_H J^2 + (2+n_H)J'^2}{J'^2} \frac{J^2 \Gamma}{\delta \omega^2},\\
\mathcal{W}_\textrm{f}/\delta \omega &= \frac{n_H (2+n_H)}{2 + 5n_H+3n_H^2 } \frac{J^2\Gamma}{\delta \omega^2},  \\
\mathcal{W}_\textrm{r}/\delta \omega &= - n_H \frac{J^2 \Gamma}{\delta \omega^2} .
\end{align*}
The rectification can be found to be
\begin{equation}
\mathcal{R} = \frac{8 n_H (2+n_H)}{2+5n_H+3n_H^2} \frac{J'^2}{8n_H J^2 + (2+n_H)J'^2} \frac{\delta \omega^2 }{ \Gamma^2}. \label{eq:app_rec}
\end{equation}
This approximate expression for the rectification and the full solution is plotted in Fig.~\ref{figure2}(d) for different values of $\Gamma$. There is a clear overlap between the two solutions, and we see that $\mathcal{R} \propto \delta \omega^2$. Furthermore, in the limit $\delta \omega \rightarrow \infty$, this model approaches the idealized model in Figs.~\ref{figure1}(a)-(b), and we achieve an ideal diode. The work both in forward and reverse bias is suppressed as $1/\delta \omega^{2}$, and therefore, the work done on the system is small. This is also verified from Fig.~\ref{figure2}(a). Thus the work done acts as a catalyst, and it does not contribute excitations so as to keep $\mathcal{J}_R \approx - \mathcal{J}_L$ both in forward and revers bias, see Eq.~\eqref{eq:current}.

\begin{figure}[t]
\centering
\includegraphics[width=1 \linewidth, angle=0]{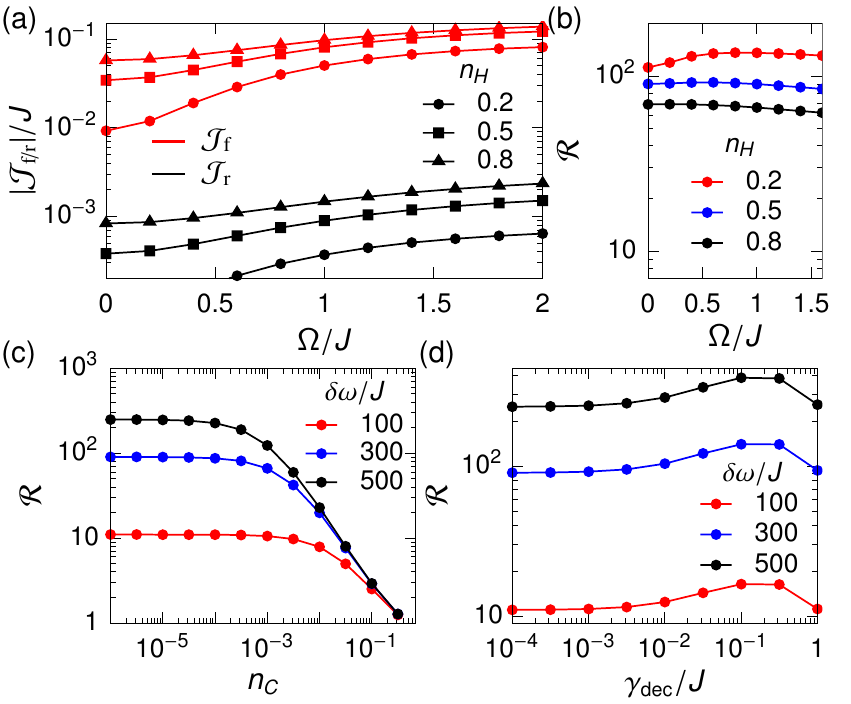}
\caption{(a) Forward-bias and reverse-bias current as a function of amplification, $\Omega$, for different values of $n_H$. (b) Rectification as a function of $\Omega$ for different values of $n_H$. (c) Rectification as a function of $n_C$ for different values of the anharmonicity, $\delta \omega$. (d) Rectification as a function of decoherence rate $\gamma_{\text{dec}}$ for different values of $\delta \omega$.}
\label{figure3}
\end{figure}

\section{Amplification} Small excitation currents can be difficult to measure, and it might be preferable to have a functioning diode with a greater heat output. Thus we might want to amplify the current through work while keeping a large rectification. This can be done by driving the transition $\ket{1_T} \leftrightarrow \ket{2_T}$ while not effecting the dark state. Since the qutrit has a large aniharmonisity, the forward-bias current can be amplified through the Hamiltonian
\begin{align*}
\hat{H}_{\text{Amp}} = \hat{H} + \frac{\Omega}{2} \left( \hat{a}_T e^{i\omega t} + \hat{a}_T^\dag e^{-i\omega t} \right).
\end{align*}
In Fig.~\ref{figure3}(a), both the forward and reverse bias currents are plotted as a function of the amplification $\Omega$. Even though the driving is only resonant in forward bias, we see an increase in both the forward-bias and reverse-bias current. For a small $n_H = 0.2$, the forward-bias current is amplified by more than an order of magnitude, while the amplification is lower for larger $n_H$. In Fig.~\ref{figure3}(b), the rectification is plotted as a function of the amplification, $\Omega$. The rectification changes very little as a function of $\Omega$.

\section{Robustness} Finally, we study the robustness towards excitations from the cold bath and decoherence. First, we let the cold bath introduce excitations by letting $n_C > 0$. The rectification is plotted in Fig.~\ref{figure3}(c) as a function of $n_C$. The diode functionality is clearly diminished for larger $n_C$. This can be explained by looking at the reverse-bias rate
\begin{align}
\Gamma_{0_T \rightarrow 1_T} = \frac{J'^2 n_C}{\Gamma} + \frac{n_H J^2 \Gamma}{\delta \omega^2}.
\end{align}
Since this process decreases the population of the dark state, it results in a decrease in rectification. Therefore, in addition to large $\delta \omega$, we need a small $n_C$. The assumption $n_C=0$ is valid when $n_C \ll n_H \frac{J^2}{J'^2} \frac{\Gamma^2}{ \delta \omega^2}$. For the default values, this corresponds to $T_C \ll 0.16\omega$, which is achievable in current quantum technology platforms \cite{doi:10.1063/1.5089550}. Second, we add decoherence in the form of decay and dephasing to the qutrit. This is done through an updated Liouvillian
\begin{align*}
\mathcal{L}_{\text{dec}}[\hat{\rho}] &= \mathcal{L}[\hat{\rho}] + \gamma_\text{dec} \left( \hat{a}_T \hat{\rho} \hat{a}_T^\dag - \frac{1}{2} \{ \hat{a}_T^\dag \hat{a}_T, \hat{\rho} \} \right) \\
& \hspace{2cm} + \gamma_\text{dec} \left( \hat{a}_T^\dag \hat{a}_T \hat{\rho} \hat{a}_T^\dag \hat{a}_T - \frac{1}{2} \{ \hat{a}_T^\dag \hat{a}_T \hat{a}_T^\dag \hat{a}_T, \hat{\rho} \} \right),
\end{align*}
where $\mathcal{L}[\hat{\rho}]$ is the Liouvillian from Eq.~\eqref{me:1}. This results in decay and dephasing coherence times of $T_1=T_2=\gamma^{-1}_\text{dec}$ for the lowest two states. On the other hand, the second excited state of the qutrit has decay coherence time $T_1 = \gamma^{-1}_\text{dec}/2$. In Fig.~\ref{figure3}(d), the rectification is plotted as a function of the decoherence, $\gamma_{\text{dec}}$. State-of-the-art quantum platforms can achieve $\gamma_\text{dec}/J < 10^{-3}$ \cite{doi:10.1146/annurev-conmatphys-031119-050605}. However, the dark-state-induced rectification is clearly not sensitive to decoherence, and other parameters can be focused on, e.g., a larger anharmonicity can be picked even if it results in larger decoherence. The stability of the rectification towards decoherence is clearly a result of the dark state being the qutrit ground state.

\section{Work to open and close diode}
\begin{figure}[t]
\centering
\includegraphics[width=1 \linewidth, angle=0]{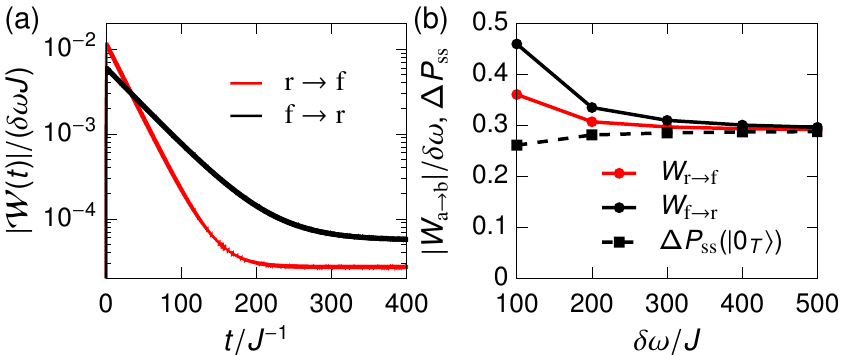}
\caption{(a) Work for an initial state of $\hat{\rho}_\mathrm{r(f)}$ and evolved with respect to the forward (reverse) bias Lindblad super operator. (b) Total work performed to transition the diode and population difference $\Delta P_\text{ss} (\ket{0_T}) = P_\textrm{ss,r}(\ket{0_T}) - P_\textrm{ss,f}(\ket{0_T})$ as a function of $\delta \omega$.}
\label{figure5}
\end{figure}

Even though approximately no work is done in steady state, the driving plays a vital role in closing and opening the diode. Therefore, work is done when the temperature bias is inverted. To study the work performed during the transition from the steady state $\hat{\rho}_\textrm{r}$ to $\hat{\rho}_\textrm{f}$, the system is prepared in $\hat{\rho}_\textrm{r}$ at $t=0$ and then evolved with respect to the forward bias Lindblad superoperator. The work done by the driving is plotted in Fig.~\ref{figure5}(a). The same is done starting in the steady state $\hat{\rho}_\textrm{f}$ and evolving with respect to the reverse bias Lindblad super operator. As expected, it requires work for the diode to switch between the steady states. Next we calculate the total energy added or subtracted through work during the transition of the diode. The total work done to transition from the reverse bias steady state to the forward bias steady state is
\begin{align*}
W_{\textrm{r}\rightarrow \textrm{f}} = \int_0^{t_\text{turn}} \mathcal{W}(t) dt
\end{align*}
where $t_\text{turn}$ is the time for the diode to go from closed to open or the reverse. We set $t_\text{turn} = 500J^{-1}$. This total work is plotted in Fig.~\ref{figure5}(b). As a comparison, we also plot
\begin{align*}
\Delta P_\text{ss} (\ket{0_T}) = P_\textrm{ss,r}(\ket{0_T}) - P_\textrm{ss,f}(\ket{0_T}).
\end{align*}
This is the population that has to be transferred for the diode to switch from closed to open. For $\delta \omega = 100J$, more work is done than is needed to make the transition possible. For larger $\delta \omega$, both $W_{\textrm{r}\rightarrow \textrm{f}}$ and $W_{\textrm{f}\rightarrow \textrm{r}}$ approach the total work needed for the transition $\Delta P_{\text{ss}}(\ket{0_T})$.

\section{Conclusion}
We have introduced a class of rectifiers that exploit the unidirectionality of a cold thermal bath to trap the system in a dark state in reverse bias. In the ideal case, the dark state is completely isolated from the hot bath and infinite rectification is achieved. Furthermore, we realized the ideal model using a single qutrit interacting with two baths mediated by two harmonic oscillators. The two harmonic oscillators transform the baths spectral densities into a sum of Lorentzians allowing for dark-state-induced rectification. We showed that rectification factors beyond $10^3$ can be achieved, and we found an approximate expression for the currents, work, and rectification. The currents can be amplified by up to an order of magnitude by driving the transition $\ket{1_T} \leftrightarrow \ket{2_T}$. We showed that the rectification is stable within achievable cold bath temperatures and towards a large degree of decoherence. Finally, we found the work needed to transition the diode from forward to reverse bias and the reverse. 

The model is simple and generic and should be realizable using several of the current quantum technology platforms like germanium quantum dots \cite{Lawrie2020}, trapped ions \cite{HAFFNER2008155}, Rydberg atoms \cite{RevModPhys.82.2313}, or superconducting circuits \cite{doi:10.1146/annurev-conmatphys-031119-050605, PRXQuantum.2.040204, doi:10.1126/science.1231930, Ronzani2018}. In superconducting circuits, the model can be implemented using a single transmon coupled capacitively to two wave guides. The waveguide correlation functions can be forced to decay using resistors \cite{Ronzani2018}. The amplification can be implemented by driving the transmon through a time-dependent flux or through capacitive driving \cite{PRXQuantum.2.040204}.

\section{Acknowledgments}

The authors acknowledge funding from The Independent Research Fund Denmark DFF-FNU. The numerical results presented in this work were obtained at the Centre for Scientific Computing, Aarhus.

\begin{figure}[t]
\centering
\includegraphics[width=1 \linewidth, angle=0]{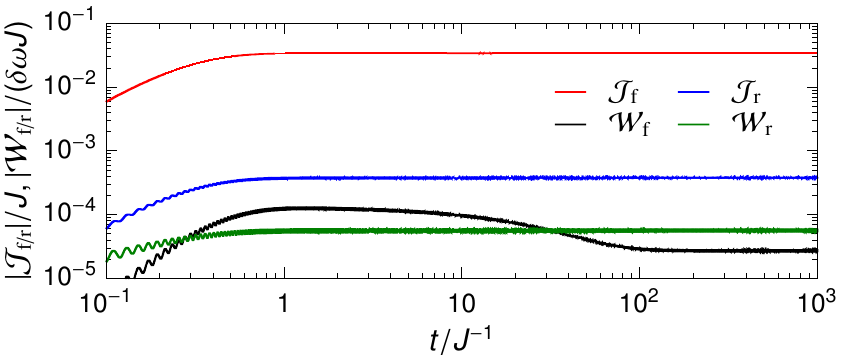}
\caption{Currents and work as a function of time in both forward and reverse bias. Here $\delta \omega = 300$, $J'=0.5J$, $\Gamma = 10J$, $n_H = 0.5$, and $n_C = 0$.}
\label{figure4}
\end{figure}

\section{Appendix A: Numerical methods and convergence of the excitation current.}

In the main text, we studied the properties of the steady-state. However, due to the time-dependent Hamiltonian the steady state will not obey $\partial_t \hat{\rho} = 0$, and it cannot be found through diagonalization. Instead, the state is evolved in time until the current and work converge. The initial state is picked to be diagonal with the approximate populations found in the main text. The harmonic oscillators are truncated such that the highest excited state $\ket{m_\text{max}}$ has $P_{\text{th}}(\ket{m_\text{max}}) < 10^{-3}$ where $m_\text{max} \in \{0, 1, 2, 3, ...\}$. Furthermore, $P_{\text{th}}(\ket{m})$ is the population of $\ket{m}$ in the harmonic-oscillator thermal state
\begin{align*}
P_\text{th}(\ket{m}) &= \frac{\tr \{ \op{m} e^{-\omega/T \hat{a}^\dag \hat{a}}\}}{\tr \{e^{-\omega/T \hat{a}^\dag \hat{a}}\}}  = \frac{n^m}{(1+n)^{m+1}}, \\ 
&\text{where} \quad n = \left( e^{\omega/T} -1 \right)^{-1}. 
\end{align*} 
The highest kept excited state can then be found to be 
\begin{align*}
m_\text{max} = \left\lceil \frac{\ln\big\{ (n+1) 10^{-3} \big\} }{\ln n - \ln (n+1)} \right\rceil
\end{align*}
where $\lceil \bullet \rceil$ is the function that returns the smallest integer greater then or equal to the input. In Fig.~\ref{figure4}, the currents and work are plotted as a function of time for both forward and reverse bias. We see that all four quantities converge for larger times. In all simulations performed for the results in the main text, the density matrix was evolved for a time $t_{\text{final}} = 5000J^{-1}$ and the quantities are averaged over times $t = 4900J^{-1} - 5000J^{-1}$.

\bibliography{bibliography}

\begin{thebibliography}{47}%
\makeatletter
\providecommand \@ifxundefined [1]{%
 \@ifx{#1\undefined}
}%
\providecommand \@ifnum [1]{%
 \ifnum #1\expandafter \@firstoftwo
 \else \expandafter \@secondoftwo
 \fi
}%
\providecommand \@ifx [1]{%
 \ifx #1\expandafter \@firstoftwo
 \else \expandafter \@secondoftwo
 \fi
}%
\providecommand \natexlab [1]{#1}%
\providecommand \enquote  [1]{``#1''}%
\providecommand \bibnamefont  [1]{#1}%
\providecommand \bibfnamefont [1]{#1}%
\providecommand \citenamefont [1]{#1}%
\providecommand \href@noop [0]{\@secondoftwo}%
\providecommand \href [0]{\begingroup \@sanitize@url \@href}%
\providecommand \@href[1]{\@@startlink{#1}\@@href}%
\providecommand \@@href[1]{\endgroup#1\@@endlink}%
\providecommand \@sanitize@url [0]{\catcode `\\12\catcode `\$12\catcode
  `\&12\catcode `\#12\catcode `\^12\catcode `\_12\catcode `\%12\relax}%
\providecommand \@@startlink[1]{}%
\providecommand \@@endlink[0]{}%
\providecommand \url  [0]{\begingroup\@sanitize@url \@url }%
\providecommand \@url [1]{\endgroup\@href {#1}{\urlprefix }}%
\providecommand \urlprefix  [0]{URL }%
\providecommand \Eprint [0]{\href }%
\providecommand \doibase [0]{https://doi.org/}%
\providecommand \selectlanguage [0]{\@gobble}%
\providecommand \bibinfo  [0]{\@secondoftwo}%
\providecommand \bibfield  [0]{\@secondoftwo}%
\providecommand \translation [1]{[#1]}%
\providecommand \BibitemOpen [0]{}%
\providecommand \bibitemStop [0]{}%
\providecommand \bibitemNoStop [0]{.\EOS\space}%
\providecommand \EOS [0]{\spacefactor3000\relax}%
\providecommand \BibitemShut  [1]{\csname bibitem#1\endcsname}%
\let\auto@bib@innerbib\@empty
\bibitem [{\citenamefont {Joulain}\ \emph {et~al.}(2016)\citenamefont
  {Joulain}, \citenamefont {Drevillon}, \citenamefont {Ezzahri},\ and\
  \citenamefont {Ordonez-Miranda}}]{PhysRevLett.116.200601}%
  \BibitemOpen
  \bibfield  {author} {\bibinfo {author} {\bibfnamefont {K.}~\bibnamefont
  {Joulain}}, \bibinfo {author} {\bibfnamefont {J.}~\bibnamefont {Drevillon}},
  \bibinfo {author} {\bibfnamefont {Y.}~\bibnamefont {Ezzahri}},\ and\ \bibinfo
  {author} {\bibfnamefont {J.}~\bibnamefont {Ordonez-Miranda}},\ }\bibfield
  {title} {\bibinfo {title} {Quantum thermal transistor},\ }\href
  {https://doi.org/10.1103/PhysRevLett.116.200601} {\bibfield  {journal}
  {\bibinfo  {journal} {Phys. Rev. Lett.}\ }\textbf {\bibinfo {volume} {116}},\
  \bibinfo {pages} {200601} (\bibinfo {year} {2016})}\BibitemShut {NoStop}%
\bibitem [{\citenamefont {Majland}\ \emph {et~al.}(2020)\citenamefont
  {Majland}, \citenamefont {Christensen},\ and\ \citenamefont
  {Zinner}}]{PhysRevB.101.184510}%
  \BibitemOpen
  \bibfield  {author} {\bibinfo {author} {\bibfnamefont {M.}~\bibnamefont
  {Majland}}, \bibinfo {author} {\bibfnamefont {K.~S.}\ \bibnamefont
  {Christensen}},\ and\ \bibinfo {author} {\bibfnamefont {N.~T.}\ \bibnamefont
  {Zinner}},\ }\bibfield  {title} {\bibinfo {title} {Quantum thermal transistor
  in superconducting circuits},\ }\href
  {https://doi.org/10.1103/PhysRevB.101.184510} {\bibfield  {journal} {\bibinfo
   {journal} {Phys. Rev. B}\ }\textbf {\bibinfo {volume} {101}},\ \bibinfo
  {pages} {184510} (\bibinfo {year} {2020})}\BibitemShut {NoStop}%
\bibitem [{\citenamefont {Guo}\ \emph {et~al.}(2018)\citenamefont {Guo},
  \citenamefont {Liu},\ and\ \citenamefont {Yu}}]{PhysRevE.98.022118}%
  \BibitemOpen
  \bibfield  {author} {\bibinfo {author} {\bibfnamefont {B.-q.}\ \bibnamefont
  {Guo}}, \bibinfo {author} {\bibfnamefont {T.}~\bibnamefont {Liu}},\ and\
  \bibinfo {author} {\bibfnamefont {C.-s.}\ \bibnamefont {Yu}},\ }\bibfield
  {title} {\bibinfo {title} {Quantum thermal transistor based on qubit-qutrit
  coupling},\ }\href {https://doi.org/10.1103/PhysRevE.98.022118} {\bibfield
  {journal} {\bibinfo  {journal} {Phys. Rev. E}\ }\textbf {\bibinfo {volume}
  {98}},\ \bibinfo {pages} {022118} (\bibinfo {year} {2018})}\BibitemShut
  {NoStop}%
\bibitem [{\citenamefont {Li}\ \emph {et~al.}(2012)\citenamefont {Li},
  \citenamefont {Ren}, \citenamefont {Wang}, \citenamefont {Zhang},
  \citenamefont {H\"anggi},\ and\ \citenamefont {Li}}]{RevModPhys.84.1045}%
  \BibitemOpen
  \bibfield  {author} {\bibinfo {author} {\bibfnamefont {N.}~\bibnamefont
  {Li}}, \bibinfo {author} {\bibfnamefont {J.}~\bibnamefont {Ren}}, \bibinfo
  {author} {\bibfnamefont {L.}~\bibnamefont {Wang}}, \bibinfo {author}
  {\bibfnamefont {G.}~\bibnamefont {Zhang}}, \bibinfo {author} {\bibfnamefont
  {P.}~\bibnamefont {H\"anggi}},\ and\ \bibinfo {author} {\bibfnamefont
  {B.}~\bibnamefont {Li}},\ }\bibfield  {title} {\bibinfo {title} {Colloquium:
  Phononics: Manipulating heat flow with electronic analogs and beyond},\
  }\href {https://doi.org/10.1103/RevModPhys.84.1045} {\bibfield  {journal}
  {\bibinfo  {journal} {Rev. Mod. Phys.}\ }\textbf {\bibinfo {volume} {84}},\
  \bibinfo {pages} {1045} (\bibinfo {year} {2012})}\BibitemShut {NoStop}%
\bibitem [{\citenamefont {Marcos-Vicioso}\ \emph {et~al.}(2018)\citenamefont
  {Marcos-Vicioso}, \citenamefont {L\'opez-Jurado}, \citenamefont
  {Ruiz-Garcia},\ and\ \citenamefont {S\'anchez}}]{PhysRevB.98.035414}%
  \BibitemOpen
  \bibfield  {author} {\bibinfo {author} {\bibfnamefont {A.}~\bibnamefont
  {Marcos-Vicioso}}, \bibinfo {author} {\bibfnamefont {C.}~\bibnamefont
  {L\'opez-Jurado}}, \bibinfo {author} {\bibfnamefont {M.}~\bibnamefont
  {Ruiz-Garcia}},\ and\ \bibinfo {author} {\bibfnamefont {R.}~\bibnamefont
  {S\'anchez}},\ }\bibfield  {title} {\bibinfo {title} {Thermal rectification
  with interacting electronic channels: Exploiting degeneracy, quantum
  superpositions, and interference},\ }\href
  {https://doi.org/10.1103/PhysRevB.98.035414} {\bibfield  {journal} {\bibinfo
  {journal} {Phys. Rev. B}\ }\textbf {\bibinfo {volume} {98}},\ \bibinfo
  {pages} {035414} (\bibinfo {year} {2018})}\BibitemShut {NoStop}%
\bibitem [{\citenamefont {Tesser}\ \emph {et~al.}(2022)\citenamefont {Tesser},
  \citenamefont {Bhandari}, \citenamefont {Erdman}, \citenamefont {Paladino},
  \citenamefont {Fazio},\ and\ \citenamefont {Taddei}}]{Tesser_2022}%
  \BibitemOpen
  \bibfield  {author} {\bibinfo {author} {\bibfnamefont {L.}~\bibnamefont
  {Tesser}}, \bibinfo {author} {\bibfnamefont {B.}~\bibnamefont {Bhandari}},
  \bibinfo {author} {\bibfnamefont {P.~A.}\ \bibnamefont {Erdman}}, \bibinfo
  {author} {\bibfnamefont {E.}~\bibnamefont {Paladino}}, \bibinfo {author}
  {\bibfnamefont {R.}~\bibnamefont {Fazio}},\ and\ \bibinfo {author}
  {\bibfnamefont {F.}~\bibnamefont {Taddei}},\ }\bibfield  {title} {\bibinfo
  {title} {Heat rectification through single and coupled quantum dots},\ }\href
  {https://doi.org/10.1088/1367-2630/ac53b8} {\bibfield  {journal} {\bibinfo
  {journal} {New Journal of Physics}\ }\textbf {\bibinfo {volume} {24}},\
  \bibinfo {pages} {035001} (\bibinfo {year} {2022})}\BibitemShut {NoStop}%
\bibitem [{\citenamefont {Peterson}\ \emph {et~al.}(2019)\citenamefont
  {Peterson}, \citenamefont {Batalh\~ao}, \citenamefont {Herrera},
  \citenamefont {Souza}, \citenamefont {Sarthour}, \citenamefont {Oliveira},\
  and\ \citenamefont {Serra}}]{PhysRevLett.123.240601}%
  \BibitemOpen
  \bibfield  {author} {\bibinfo {author} {\bibfnamefont {J.~P.~S.}\
  \bibnamefont {Peterson}}, \bibinfo {author} {\bibfnamefont {T.~B.}\
  \bibnamefont {Batalh\~ao}}, \bibinfo {author} {\bibfnamefont
  {M.}~\bibnamefont {Herrera}}, \bibinfo {author} {\bibfnamefont {A.~M.}\
  \bibnamefont {Souza}}, \bibinfo {author} {\bibfnamefont {R.~S.}\ \bibnamefont
  {Sarthour}}, \bibinfo {author} {\bibfnamefont {I.~S.}\ \bibnamefont
  {Oliveira}},\ and\ \bibinfo {author} {\bibfnamefont {R.~M.}\ \bibnamefont
  {Serra}},\ }\bibfield  {title} {\bibinfo {title} {Experimental
  characterization of a spin quantum heat engine},\ }\href
  {https://doi.org/10.1103/PhysRevLett.123.240601} {\bibfield  {journal}
  {\bibinfo  {journal} {Phys. Rev. Lett.}\ }\textbf {\bibinfo {volume} {123}},\
  \bibinfo {pages} {240601} (\bibinfo {year} {2019})}\BibitemShut {NoStop}%
\bibitem [{\citenamefont {Ono}\ \emph {et~al.}(2020)\citenamefont {Ono},
  \citenamefont {Shevchenko}, \citenamefont {Mori}, \citenamefont {Moriyama},\
  and\ \citenamefont {Nori}}]{PhysRevLett.125.166802}%
  \BibitemOpen
  \bibfield  {author} {\bibinfo {author} {\bibfnamefont {K.}~\bibnamefont
  {Ono}}, \bibinfo {author} {\bibfnamefont {S.~N.}\ \bibnamefont {Shevchenko}},
  \bibinfo {author} {\bibfnamefont {T.}~\bibnamefont {Mori}}, \bibinfo {author}
  {\bibfnamefont {S.}~\bibnamefont {Moriyama}},\ and\ \bibinfo {author}
  {\bibfnamefont {F.}~\bibnamefont {Nori}},\ }\bibfield  {title} {\bibinfo
  {title} {Analog of a quantum heat engine using a single-spin qubit},\ }\href
  {https://doi.org/10.1103/PhysRevLett.125.166802} {\bibfield  {journal}
  {\bibinfo  {journal} {Phys. Rev. Lett.}\ }\textbf {\bibinfo {volume} {125}},\
  \bibinfo {pages} {166802} (\bibinfo {year} {2020})}\BibitemShut {NoStop}%
\bibitem [{\citenamefont {Josefsson}\ \emph {et~al.}(2018)\citenamefont
  {Josefsson}, \citenamefont {Svilans}, \citenamefont {Burke}, \citenamefont
  {Hoffmann}, \citenamefont {Fahlvik}, \citenamefont {Thelander}, \citenamefont
  {Leijnse},\ and\ \citenamefont {Linke}}]{Josefsson2018}%
  \BibitemOpen
  \bibfield  {author} {\bibinfo {author} {\bibfnamefont {M.}~\bibnamefont
  {Josefsson}}, \bibinfo {author} {\bibfnamefont {A.}~\bibnamefont {Svilans}},
  \bibinfo {author} {\bibfnamefont {A.~M.}\ \bibnamefont {Burke}}, \bibinfo
  {author} {\bibfnamefont {E.~A.}\ \bibnamefont {Hoffmann}}, \bibinfo {author}
  {\bibfnamefont {S.}~\bibnamefont {Fahlvik}}, \bibinfo {author} {\bibfnamefont
  {C.}~\bibnamefont {Thelander}}, \bibinfo {author} {\bibfnamefont
  {M.}~\bibnamefont {Leijnse}},\ and\ \bibinfo {author} {\bibfnamefont
  {H.}~\bibnamefont {Linke}},\ }\bibfield  {title} {\bibinfo {title} {A
  quantum-dot heat engine operating close to the thermodynamic efficiency
  limits},\ }\href {https://doi.org/10.1038/s41565-018-0200-5} {\bibfield
  {journal} {\bibinfo  {journal} {Nat. Nanotechnol.}\ }\textbf {\bibinfo
  {volume} {13}},\ \bibinfo {pages} {920} (\bibinfo {year} {2018})}\BibitemShut
  {NoStop}%
\bibitem [{\citenamefont {Koski}\ \emph {et~al.}(2014)\citenamefont {Koski},
  \citenamefont {Maisi}, \citenamefont {Sagawa},\ and\ \citenamefont
  {Pekola}}]{PhysRevLett.113.030601}%
  \BibitemOpen
  \bibfield  {author} {\bibinfo {author} {\bibfnamefont {J.~V.}\ \bibnamefont
  {Koski}}, \bibinfo {author} {\bibfnamefont {V.~F.}\ \bibnamefont {Maisi}},
  \bibinfo {author} {\bibfnamefont {T.}~\bibnamefont {Sagawa}},\ and\ \bibinfo
  {author} {\bibfnamefont {J.~P.}\ \bibnamefont {Pekola}},\ }\bibfield  {title}
  {\bibinfo {title} {Experimental observation of the role of mutual information
  in the nonequilibrium dynamics of a {M}axwell demon},\ }\href
  {https://doi.org/10.1103/PhysRevLett.113.030601} {\bibfield  {journal}
  {\bibinfo  {journal} {Phys. Rev. Lett.}\ }\textbf {\bibinfo {volume} {113}},\
  \bibinfo {pages} {030601} (\bibinfo {year} {2014})}\BibitemShut {NoStop}%
\bibitem [{\citenamefont {Naghiloo}\ \emph {et~al.}(2018)\citenamefont
  {Naghiloo}, \citenamefont {Alonso}, \citenamefont {Romito}, \citenamefont
  {Lutz},\ and\ \citenamefont {Murch}}]{PhysRevLett.121.030604}%
  \BibitemOpen
  \bibfield  {author} {\bibinfo {author} {\bibfnamefont {M.}~\bibnamefont
  {Naghiloo}}, \bibinfo {author} {\bibfnamefont {J.~J.}\ \bibnamefont
  {Alonso}}, \bibinfo {author} {\bibfnamefont {A.}~\bibnamefont {Romito}},
  \bibinfo {author} {\bibfnamefont {E.}~\bibnamefont {Lutz}},\ and\ \bibinfo
  {author} {\bibfnamefont {K.~W.}\ \bibnamefont {Murch}},\ }\bibfield  {title}
  {\bibinfo {title} {Information gain and loss for a quantum {M}axwell's
  demon},\ }\href {https://doi.org/10.1103/PhysRevLett.121.030604} {\bibfield
  {journal} {\bibinfo  {journal} {Phys. Rev. Lett.}\ }\textbf {\bibinfo
  {volume} {121}},\ \bibinfo {pages} {030604} (\bibinfo {year}
  {2018})}\BibitemShut {NoStop}%
\bibitem [{\citenamefont {Poulsen}\ \emph
  {et~al.}(2022{\natexlab{a}})\citenamefont {Poulsen}, \citenamefont {Majland},
  \citenamefont {Lloyd}, \citenamefont {Kjaergaard},\ and\ \citenamefont
  {Zinner}}]{poulsen2021non}%
  \BibitemOpen
  \bibfield  {author} {\bibinfo {author} {\bibfnamefont {K.}~\bibnamefont
  {Poulsen}}, \bibinfo {author} {\bibfnamefont {M.}~\bibnamefont {Majland}},
  \bibinfo {author} {\bibfnamefont {S.}~\bibnamefont {Lloyd}}, \bibinfo
  {author} {\bibfnamefont {M.}~\bibnamefont {Kjaergaard}},\ and\ \bibinfo
  {author} {\bibfnamefont {N.~T.}\ \bibnamefont {Zinner}},\ }\bibfield  {title}
  {\bibinfo {title} {Quantum maxwell's demon assisted by non-markovian
  effects},\ }\href {https://doi.org/10.1103/PhysRevE.105.044141} {\bibfield
  {journal} {\bibinfo  {journal} {Phys. Rev. E}\ }\textbf {\bibinfo {volume}
  {105}},\ \bibinfo {pages} {044141} (\bibinfo {year}
  {2022}{\natexlab{a}})}\BibitemShut {NoStop}%
\bibitem [{\citenamefont {Landi}\ \emph {et~al.}(2021)\citenamefont {Landi},
  \citenamefont {Poletti},\ and\ \citenamefont {Schaller}}]{landi2021non}%
  \BibitemOpen
  \bibfield  {author} {\bibinfo {author} {\bibfnamefont {G.~T.}\ \bibnamefont
  {Landi}}, \bibinfo {author} {\bibfnamefont {D.}~\bibnamefont {Poletti}},\
  and\ \bibinfo {author} {\bibfnamefont {G.}~\bibnamefont {Schaller}},\
  }\bibfield  {title} {\bibinfo {title} {Non-equilibrium boundary driven
  quantum systems: models, methods and properties},\ }\href@noop {} {\bibfield
  {journal} {\bibinfo  {journal} {arXiv preprint arXiv:2104.14350}\ } (\bibinfo
  {year} {2021})}\BibitemShut {NoStop}%
\bibitem [{\citenamefont {Prosen}(2011)}]{PhysRevLett.106.217206}%
  \BibitemOpen
  \bibfield  {author} {\bibinfo {author} {\bibfnamefont {T.}~\bibnamefont
  {Prosen}},\ }\bibfield  {title} {\bibinfo {title} {Open $xxz$ spin chain:
  Nonequilibrium steady state and a strict bound on ballistic transport},\
  }\href {https://doi.org/10.1103/PhysRevLett.106.217206} {\bibfield  {journal}
  {\bibinfo  {journal} {Phys. Rev. Lett.}\ }\textbf {\bibinfo {volume} {106}},\
  \bibinfo {pages} {217206} (\bibinfo {year} {2011})}\BibitemShut {NoStop}%
\bibitem [{\citenamefont {Poulsen}\ and\ \citenamefont
  {Zinner}(2021)}]{PhysRevLett.126.077203}%
  \BibitemOpen
  \bibfield  {author} {\bibinfo {author} {\bibfnamefont {K.}~\bibnamefont
  {Poulsen}}\ and\ \bibinfo {author} {\bibfnamefont {N.~T.}\ \bibnamefont
  {Zinner}},\ }\bibfield  {title} {\bibinfo {title} {Giant magnetoresistance in
  boundary-driven spin chains},\ }\href
  {https://doi.org/10.1103/PhysRevLett.126.077203} {\bibfield  {journal}
  {\bibinfo  {journal} {Phys. Rev. Lett.}\ }\textbf {\bibinfo {volume} {126}},\
  \bibinfo {pages} {077203} (\bibinfo {year} {2021})}\BibitemShut {NoStop}%
\bibitem [{\citenamefont {Segal}(2006)}]{PhysRevB.73.205415}%
  \BibitemOpen
  \bibfield  {author} {\bibinfo {author} {\bibfnamefont {D.}~\bibnamefont
  {Segal}},\ }\bibfield  {title} {\bibinfo {title} {Heat flow in nonlinear
  molecular junctions: Master equation analysis},\ }\href
  {https://doi.org/10.1103/PhysRevB.73.205415} {\bibfield  {journal} {\bibinfo
  {journal} {Phys. Rev. B}\ }\textbf {\bibinfo {volume} {73}},\ \bibinfo
  {pages} {205415} (\bibinfo {year} {2006})}\BibitemShut {NoStop}%
\bibitem [{\citenamefont {Senior}\ \emph {et~al.}(2020)\citenamefont {Senior},
  \citenamefont {Gubaydullin}, \citenamefont {Karimi}, \citenamefont
  {Peltonen}, \citenamefont {Ankerhold},\ and\ \citenamefont
  {Pekola}}]{Senior2020}%
  \BibitemOpen
  \bibfield  {author} {\bibinfo {author} {\bibfnamefont {J.}~\bibnamefont
  {Senior}}, \bibinfo {author} {\bibfnamefont {A.}~\bibnamefont {Gubaydullin}},
  \bibinfo {author} {\bibfnamefont {B.}~\bibnamefont {Karimi}}, \bibinfo
  {author} {\bibfnamefont {J.~T.}\ \bibnamefont {Peltonen}}, \bibinfo {author}
  {\bibfnamefont {J.}~\bibnamefont {Ankerhold}},\ and\ \bibinfo {author}
  {\bibfnamefont {J.~P.}\ \bibnamefont {Pekola}},\ }\bibfield  {title}
  {\bibinfo {title} {Heat rectification via a superconducting artificial
  atom},\ }\href {https://doi.org/10.1038/s42005-020-0307-5} {\bibfield
  {journal} {\bibinfo  {journal} {Commun. Phys.}\ }\textbf {\bibinfo {volume}
  {3}},\ \bibinfo {pages} {40} (\bibinfo {year} {2020})}\BibitemShut {NoStop}%
\bibitem [{\citenamefont {Bhandari}\ \emph {et~al.}(2021)\citenamefont
  {Bhandari}, \citenamefont {Erdman}, \citenamefont {Fazio}, \citenamefont
  {Paladino},\ and\ \citenamefont {Taddei}}]{PhysRevB.103.155434}%
  \BibitemOpen
  \bibfield  {author} {\bibinfo {author} {\bibfnamefont {B.}~\bibnamefont
  {Bhandari}}, \bibinfo {author} {\bibfnamefont {P.~A.}\ \bibnamefont
  {Erdman}}, \bibinfo {author} {\bibfnamefont {R.}~\bibnamefont {Fazio}},
  \bibinfo {author} {\bibfnamefont {E.}~\bibnamefont {Paladino}},\ and\
  \bibinfo {author} {\bibfnamefont {F.}~\bibnamefont {Taddei}},\ }\bibfield
  {title} {\bibinfo {title} {Thermal rectification through a nonlinear quantum
  resonator},\ }\href {https://doi.org/10.1103/PhysRevB.103.155434} {\bibfield
  {journal} {\bibinfo  {journal} {Phys. Rev. B}\ }\textbf {\bibinfo {volume}
  {103}},\ \bibinfo {pages} {155434} (\bibinfo {year} {2021})}\BibitemShut
  {NoStop}%
\bibitem [{\citenamefont {Werlang}\ \emph {et~al.}(2014)\citenamefont
  {Werlang}, \citenamefont {Marchiori}, \citenamefont {Cornelio},\ and\
  \citenamefont {Valente}}]{PhysRevE.89.062109}%
  \BibitemOpen
  \bibfield  {author} {\bibinfo {author} {\bibfnamefont {T.}~\bibnamefont
  {Werlang}}, \bibinfo {author} {\bibfnamefont {M.~A.}\ \bibnamefont
  {Marchiori}}, \bibinfo {author} {\bibfnamefont {M.~F.}\ \bibnamefont
  {Cornelio}},\ and\ \bibinfo {author} {\bibfnamefont {D.}~\bibnamefont
  {Valente}},\ }\bibfield  {title} {\bibinfo {title} {Optimal rectification in
  the ultrastrong coupling regime},\ }\href
  {https://doi.org/10.1103/PhysRevE.89.062109} {\bibfield  {journal} {\bibinfo
  {journal} {Phys. Rev. E}\ }\textbf {\bibinfo {volume} {89}},\ \bibinfo
  {pages} {062109} (\bibinfo {year} {2014})}\BibitemShut {NoStop}%
\bibitem [{\citenamefont {Iorio}\ \emph {et~al.}(2021)\citenamefont {Iorio},
  \citenamefont {Strambini}, \citenamefont {Haack}, \citenamefont {Campisi},\
  and\ \citenamefont {Giazotto}}]{PhysRevApplied.15.054050}%
  \BibitemOpen
  \bibfield  {author} {\bibinfo {author} {\bibfnamefont {A.}~\bibnamefont
  {Iorio}}, \bibinfo {author} {\bibfnamefont {E.}~\bibnamefont {Strambini}},
  \bibinfo {author} {\bibfnamefont {G.}~\bibnamefont {Haack}}, \bibinfo
  {author} {\bibfnamefont {M.}~\bibnamefont {Campisi}},\ and\ \bibinfo {author}
  {\bibfnamefont {F.}~\bibnamefont {Giazotto}},\ }\bibfield  {title} {\bibinfo
  {title} {Photonic heat rectification in a system of coupled qubits},\ }\href
  {https://doi.org/10.1103/PhysRevApplied.15.054050} {\bibfield  {journal}
  {\bibinfo  {journal} {Phys. Rev. Appl.}\ }\textbf {\bibinfo {volume} {15}},\
  \bibinfo {pages} {054050} (\bibinfo {year} {2021})}\BibitemShut {NoStop}%
\bibitem [{\citenamefont {Yan}\ \emph {et~al.}(2009)\citenamefont {Yan},
  \citenamefont {Wu},\ and\ \citenamefont {Li}}]{PhysRevB.79.014207}%
  \BibitemOpen
  \bibfield  {author} {\bibinfo {author} {\bibfnamefont {Y.}~\bibnamefont
  {Yan}}, \bibinfo {author} {\bibfnamefont {C.-Q.}\ \bibnamefont {Wu}},\ and\
  \bibinfo {author} {\bibfnamefont {B.}~\bibnamefont {Li}},\ }\bibfield
  {title} {\bibinfo {title} {Control of heat transport in quantum spin
  systems},\ }\href {https://doi.org/10.1103/PhysRevB.79.014207} {\bibfield
  {journal} {\bibinfo  {journal} {Phys. Rev. B}\ }\textbf {\bibinfo {volume}
  {79}},\ \bibinfo {pages} {014207} (\bibinfo {year} {2009})}\BibitemShut
  {NoStop}%
\bibitem [{\citenamefont {Balachandran}\ \emph {et~al.}(2018)\citenamefont
  {Balachandran}, \citenamefont {Benenti}, \citenamefont {Pereira},
  \citenamefont {Casati},\ and\ \citenamefont
  {Poletti}}]{PhysRevLett.120.200603}%
  \BibitemOpen
  \bibfield  {author} {\bibinfo {author} {\bibfnamefont {V.}~\bibnamefont
  {Balachandran}}, \bibinfo {author} {\bibfnamefont {G.}~\bibnamefont
  {Benenti}}, \bibinfo {author} {\bibfnamefont {E.}~\bibnamefont {Pereira}},
  \bibinfo {author} {\bibfnamefont {G.}~\bibnamefont {Casati}},\ and\ \bibinfo
  {author} {\bibfnamefont {D.}~\bibnamefont {Poletti}},\ }\bibfield  {title}
  {\bibinfo {title} {Perfect diode in quantum spin chains},\ }\href
  {https://doi.org/10.1103/PhysRevLett.120.200603} {\bibfield  {journal}
  {\bibinfo  {journal} {Phys. Rev. Lett.}\ }\textbf {\bibinfo {volume} {120}},\
  \bibinfo {pages} {200603} (\bibinfo {year} {2018})}\BibitemShut {NoStop}%
\bibitem [{\citenamefont {Zhang}\ \emph {et~al.}(2009)\citenamefont {Zhang},
  \citenamefont {Yan}, \citenamefont {Wu}, \citenamefont {Wang},\ and\
  \citenamefont {Li}}]{PhysRevB.80.172301}%
  \BibitemOpen
  \bibfield  {author} {\bibinfo {author} {\bibfnamefont {L.}~\bibnamefont
  {Zhang}}, \bibinfo {author} {\bibfnamefont {Y.}~\bibnamefont {Yan}}, \bibinfo
  {author} {\bibfnamefont {C.-Q.}\ \bibnamefont {Wu}}, \bibinfo {author}
  {\bibfnamefont {J.-S.}\ \bibnamefont {Wang}},\ and\ \bibinfo {author}
  {\bibfnamefont {B.}~\bibnamefont {Li}},\ }\bibfield  {title} {\bibinfo
  {title} {Reversal of thermal rectification in quantum systems},\ }\href
  {https://doi.org/10.1103/PhysRevB.80.172301} {\bibfield  {journal} {\bibinfo
  {journal} {Phys. Rev. B}\ }\textbf {\bibinfo {volume} {80}},\ \bibinfo
  {pages} {172301} (\bibinfo {year} {2009})}\BibitemShut {NoStop}%
\bibitem [{\citenamefont {Balachandran}\ \emph {et~al.}(2019)\citenamefont
  {Balachandran}, \citenamefont {Benenti}, \citenamefont {Pereira},
  \citenamefont {Casati},\ and\ \citenamefont {Poletti}}]{PhysRevE.99.032136}%
  \BibitemOpen
  \bibfield  {author} {\bibinfo {author} {\bibfnamefont {V.}~\bibnamefont
  {Balachandran}}, \bibinfo {author} {\bibfnamefont {G.}~\bibnamefont
  {Benenti}}, \bibinfo {author} {\bibfnamefont {E.}~\bibnamefont {Pereira}},
  \bibinfo {author} {\bibfnamefont {G.}~\bibnamefont {Casati}},\ and\ \bibinfo
  {author} {\bibfnamefont {D.}~\bibnamefont {Poletti}},\ }\bibfield  {title}
  {\bibinfo {title} {Heat current rectification in segmented $xxz$ chains},\
  }\href {https://doi.org/10.1103/PhysRevE.99.032136} {\bibfield  {journal}
  {\bibinfo  {journal} {Phys. Rev. E}\ }\textbf {\bibinfo {volume} {99}},\
  \bibinfo {pages} {032136} (\bibinfo {year} {2019})}\BibitemShut {NoStop}%
\bibitem [{\citenamefont {Silva}\ \emph {et~al.}(2020)\citenamefont {Silva},
  \citenamefont {Landi}, \citenamefont {Drumond},\ and\ \citenamefont
  {Pereira}}]{PhysRevE.102.062146}%
  \BibitemOpen
  \bibfield  {author} {\bibinfo {author} {\bibfnamefont {S.~H.~S.}\
  \bibnamefont {Silva}}, \bibinfo {author} {\bibfnamefont {G.~T.}\ \bibnamefont
  {Landi}}, \bibinfo {author} {\bibfnamefont {R.~C.}\ \bibnamefont {Drumond}},\
  and\ \bibinfo {author} {\bibfnamefont {E.}~\bibnamefont {Pereira}},\
  }\bibfield  {title} {\bibinfo {title} {Heat rectification on the $xx$
  chain},\ }\href {https://doi.org/10.1103/PhysRevE.102.062146} {\bibfield
  {journal} {\bibinfo  {journal} {Phys. Rev. E}\ }\textbf {\bibinfo {volume}
  {102}},\ \bibinfo {pages} {062146} (\bibinfo {year} {2020})}\BibitemShut
  {NoStop}%
\bibitem [{\citenamefont {Chioquetta}\ \emph {et~al.}(2021)\citenamefont
  {Chioquetta}, \citenamefont {Pereira}, \citenamefont {Landi},\ and\
  \citenamefont {Drumond}}]{PhysRevE.103.032108}%
  \BibitemOpen
  \bibfield  {author} {\bibinfo {author} {\bibfnamefont {A.}~\bibnamefont
  {Chioquetta}}, \bibinfo {author} {\bibfnamefont {E.}~\bibnamefont {Pereira}},
  \bibinfo {author} {\bibfnamefont {G.~T.}\ \bibnamefont {Landi}},\ and\
  \bibinfo {author} {\bibfnamefont {R.~C.}\ \bibnamefont {Drumond}},\
  }\bibfield  {title} {\bibinfo {title} {Rectification induced by geometry in
  two-dimensional quantum spin lattices},\ }\href
  {https://doi.org/10.1103/PhysRevE.103.032108} {\bibfield  {journal} {\bibinfo
   {journal} {Phys. Rev. E}\ }\textbf {\bibinfo {volume} {103}},\ \bibinfo
  {pages} {032108} (\bibinfo {year} {2021})}\BibitemShut {NoStop}%
\bibitem [{\citenamefont {Poulsen}\ \emph
  {et~al.}(2022{\natexlab{b}})\citenamefont {Poulsen}, \citenamefont {Santos},
  \citenamefont {Kristensen},\ and\ \citenamefont
  {Zinner}}]{poulsen2021entanglement}%
  \BibitemOpen
  \bibfield  {author} {\bibinfo {author} {\bibfnamefont {K.}~\bibnamefont
  {Poulsen}}, \bibinfo {author} {\bibfnamefont {A.~C.}\ \bibnamefont {Santos}},
  \bibinfo {author} {\bibfnamefont {L.~B.}\ \bibnamefont {Kristensen}},\ and\
  \bibinfo {author} {\bibfnamefont {N.~T.}\ \bibnamefont {Zinner}},\ }\bibfield
   {title} {\bibinfo {title} {Entanglement-enhanced quantum rectification},\
  }\href {https://doi.org/10.1103/PhysRevA.105.052605} {\bibfield  {journal}
  {\bibinfo  {journal} {Phys. Rev. A}\ }\textbf {\bibinfo {volume} {105}},\
  \bibinfo {pages} {052605} (\bibinfo {year} {2022}{\natexlab{b}})}\BibitemShut
  {NoStop}%
\bibitem [{\citenamefont {Lee}\ \emph {et~al.}(2022)\citenamefont {Lee},
  \citenamefont {Balachandran}, \citenamefont {Guo},\ and\ \citenamefont
  {Poletti}}]{PhysRevE.105.024120}%
  \BibitemOpen
  \bibfield  {author} {\bibinfo {author} {\bibfnamefont {K.~H.}\ \bibnamefont
  {Lee}}, \bibinfo {author} {\bibfnamefont {V.}~\bibnamefont {Balachandran}},
  \bibinfo {author} {\bibfnamefont {C.}~\bibnamefont {Guo}},\ and\ \bibinfo
  {author} {\bibfnamefont {D.}~\bibnamefont {Poletti}},\ }\bibfield  {title}
  {\bibinfo {title} {Transport and spectral properties of the $xx+xxz$ diode
  and stability to dephasing},\ }\href
  {https://doi.org/10.1103/PhysRevE.105.024120} {\bibfield  {journal} {\bibinfo
   {journal} {Phys. Rev. E}\ }\textbf {\bibinfo {volume} {105}},\ \bibinfo
  {pages} {024120} (\bibinfo {year} {2022})}\BibitemShut {NoStop}%
\bibitem [{\citenamefont {Fleischhauer}\ \emph {et~al.}(2005)\citenamefont
  {Fleischhauer}, \citenamefont {Imamoglu},\ and\ \citenamefont
  {Marangos}}]{RevModPhys.77.633}%
  \BibitemOpen
  \bibfield  {author} {\bibinfo {author} {\bibfnamefont {M.}~\bibnamefont
  {Fleischhauer}}, \bibinfo {author} {\bibfnamefont {A.}~\bibnamefont
  {Imamoglu}},\ and\ \bibinfo {author} {\bibfnamefont {J.~P.}\ \bibnamefont
  {Marangos}},\ }\bibfield  {title} {\bibinfo {title} {Electromagnetically
  induced transparency: Optics in coherent media},\ }\href
  {https://doi.org/10.1103/RevModPhys.77.633} {\bibfield  {journal} {\bibinfo
  {journal} {Rev. Mod. Phys.}\ }\textbf {\bibinfo {volume} {77}},\ \bibinfo
  {pages} {633} (\bibinfo {year} {2005})}\BibitemShut {NoStop}%
\bibitem [{\citenamefont {Quach}\ and\ \citenamefont
  {Munro}(2020)}]{PhysRevApplied.14.024092}%
  \BibitemOpen
  \bibfield  {author} {\bibinfo {author} {\bibfnamefont {J.~Q.}\ \bibnamefont
  {Quach}}\ and\ \bibinfo {author} {\bibfnamefont {W.~J.}\ \bibnamefont
  {Munro}},\ }\bibfield  {title} {\bibinfo {title} {Using dark states to charge
  and stabilize open quantum batteries},\ }\href
  {https://doi.org/10.1103/PhysRevApplied.14.024092} {\bibfield  {journal}
  {\bibinfo  {journal} {Phys. Rev. Applied}\ }\textbf {\bibinfo {volume}
  {14}},\ \bibinfo {pages} {024092} (\bibinfo {year} {2020})}\BibitemShut
  {NoStop}%
\bibitem [{\citenamefont {Lai}\ \emph {et~al.}(2020{\natexlab{a}})\citenamefont
  {Lai}, \citenamefont {Huang}, \citenamefont {Yin}, \citenamefont {Hou},
  \citenamefont {Li}, \citenamefont {Vitali}, \citenamefont {Nori},\ and\
  \citenamefont {Liao}}]{PhysRevA.102.011502}%
  \BibitemOpen
  \bibfield  {author} {\bibinfo {author} {\bibfnamefont {D.-G.}\ \bibnamefont
  {Lai}}, \bibinfo {author} {\bibfnamefont {J.-F.}\ \bibnamefont {Huang}},
  \bibinfo {author} {\bibfnamefont {X.-L.}\ \bibnamefont {Yin}}, \bibinfo
  {author} {\bibfnamefont {B.-P.}\ \bibnamefont {Hou}}, \bibinfo {author}
  {\bibfnamefont {W.}~\bibnamefont {Li}}, \bibinfo {author} {\bibfnamefont
  {D.}~\bibnamefont {Vitali}}, \bibinfo {author} {\bibfnamefont
  {F.}~\bibnamefont {Nori}},\ and\ \bibinfo {author} {\bibfnamefont {J.-Q.}\
  \bibnamefont {Liao}},\ }\bibfield  {title} {\bibinfo {title} {Nonreciprocal
  ground-state cooling of multiple mechanical resonators},\ }\href
  {https://doi.org/10.1103/PhysRevA.102.011502} {\bibfield  {journal} {\bibinfo
   {journal} {Phys. Rev. A}\ }\textbf {\bibinfo {volume} {102}},\ \bibinfo
  {pages} {011502(R)} (\bibinfo {year} {2020}{\natexlab{a}})}\BibitemShut
  {NoStop}%
\bibitem [{\citenamefont {Lai}\ \emph {et~al.}(2020{\natexlab{b}})\citenamefont
  {Lai}, \citenamefont {Wang}, \citenamefont {Qin}, \citenamefont {Hou},
  \citenamefont {Nori},\ and\ \citenamefont {Liao}}]{PhysRevA.102.023707}%
  \BibitemOpen
  \bibfield  {author} {\bibinfo {author} {\bibfnamefont {D.-G.}\ \bibnamefont
  {Lai}}, \bibinfo {author} {\bibfnamefont {X.}~\bibnamefont {Wang}}, \bibinfo
  {author} {\bibfnamefont {W.}~\bibnamefont {Qin}}, \bibinfo {author}
  {\bibfnamefont {B.-P.}\ \bibnamefont {Hou}}, \bibinfo {author} {\bibfnamefont
  {F.}~\bibnamefont {Nori}},\ and\ \bibinfo {author} {\bibfnamefont {J.-Q.}\
  \bibnamefont {Liao}},\ }\bibfield  {title} {\bibinfo {title} {Tunable
  optomechanically induced transparency by controlling the dark-mode effect},\
  }\href {https://doi.org/10.1103/PhysRevA.102.023707} {\bibfield  {journal}
  {\bibinfo  {journal} {Phys. Rev. A}\ }\textbf {\bibinfo {volume} {102}},\
  \bibinfo {pages} {023707} (\bibinfo {year} {2020}{\natexlab{b}})}\BibitemShut
  {NoStop}%
\bibitem [{\citenamefont {D{\'{\i}}az}\ and\ \citenamefont
  {S{\'{a}}nchez}(2021)}]{D_az_2021}%
  \BibitemOpen
  \bibfield  {author} {\bibinfo {author} {\bibfnamefont {I.}~\bibnamefont
  {D{\'{\i}}az}}\ and\ \bibinfo {author} {\bibfnamefont {R.}~\bibnamefont
  {S{\'{a}}nchez}},\ }\bibfield  {title} {\bibinfo {title} {The qutrit as a
  heat diode and circulator},\ }\href
  {https://doi.org/10.1088/1367-2630/ac4211} {\bibfield  {journal} {\bibinfo
  {journal} {New J. Phys.}\ }\textbf {\bibinfo {volume} {23}},\ \bibinfo
  {pages} {125006} (\bibinfo {year} {2021})}\BibitemShut {NoStop}%
\bibitem [{\citenamefont {Krantz}\ \emph {et~al.}(2019)\citenamefont {Krantz},
  \citenamefont {Kjaergaard}, \citenamefont {Yan}, \citenamefont {Orlando},
  \citenamefont {Gustavsson},\ and\ \citenamefont
  {Oliver}}]{doi:10.1063/1.5089550}%
  \BibitemOpen
  \bibfield  {author} {\bibinfo {author} {\bibfnamefont {P.}~\bibnamefont
  {Krantz}}, \bibinfo {author} {\bibfnamefont {M.}~\bibnamefont {Kjaergaard}},
  \bibinfo {author} {\bibfnamefont {F.}~\bibnamefont {Yan}}, \bibinfo {author}
  {\bibfnamefont {T.~P.}\ \bibnamefont {Orlando}}, \bibinfo {author}
  {\bibfnamefont {S.}~\bibnamefont {Gustavsson}},\ and\ \bibinfo {author}
  {\bibfnamefont {W.~D.}\ \bibnamefont {Oliver}},\ }\bibfield  {title}
  {\bibinfo {title} {A quantum engineer's guide to superconducting qubits},\
  }\href {https://doi.org/10.1063/1.5089550} {\bibfield  {journal} {\bibinfo
  {journal} {Appl. Phys. Rev.}\ }\textbf {\bibinfo {volume} {6}},\ \bibinfo
  {pages} {021318} (\bibinfo {year} {2019})}\BibitemShut {NoStop}%
\bibitem [{\citenamefont {Najera-Santos}\ \emph {et~al.}(2020)\citenamefont
  {Najera-Santos}, \citenamefont {Camati}, \citenamefont {M\'etillon},
  \citenamefont {Brune}, \citenamefont {Raimond}, \citenamefont {Auff\`eves},\
  and\ \citenamefont {Dotsenko}}]{PhysRevResearch.2.032025}%
  \BibitemOpen
  \bibfield  {author} {\bibinfo {author} {\bibfnamefont {B.-L.}\ \bibnamefont
  {Najera-Santos}}, \bibinfo {author} {\bibfnamefont {P.~A.}\ \bibnamefont
  {Camati}}, \bibinfo {author} {\bibfnamefont {V.}~\bibnamefont {M\'etillon}},
  \bibinfo {author} {\bibfnamefont {M.}~\bibnamefont {Brune}}, \bibinfo
  {author} {\bibfnamefont {J.-M.}\ \bibnamefont {Raimond}}, \bibinfo {author}
  {\bibfnamefont {A.}~\bibnamefont {Auff\`eves}},\ and\ \bibinfo {author}
  {\bibfnamefont {I.}~\bibnamefont {Dotsenko}},\ }\bibfield  {title} {\bibinfo
  {title} {Autonomous {M}axwell's demon in a cavity qed system},\ }\href
  {https://doi.org/10.1103/PhysRevResearch.2.032025} {\bibfield  {journal}
  {\bibinfo  {journal} {Phys. Rev. Research}\ }\textbf {\bibinfo {volume}
  {2}},\ \bibinfo {pages} {032025(R)} (\bibinfo {year} {2020})}\BibitemShut
  {NoStop}%
\bibitem [{\citenamefont {Santos}\ \emph {et~al.}(2019)\citenamefont {Santos},
  \citenamefont {\ifmmode~\mbox{\c{C}}\else \c{C}\fi{}akmak}, \citenamefont
  {Campbell},\ and\ \citenamefont {Zinner}}]{PhysRevE.100.032107}%
  \BibitemOpen
  \bibfield  {author} {\bibinfo {author} {\bibfnamefont {A.~C.}\ \bibnamefont
  {Santos}}, \bibinfo {author} {\bibfnamefont {B.}~\bibnamefont
  {\ifmmode~\mbox{\c{C}}\else \c{C}\fi{}akmak}}, \bibinfo {author}
  {\bibfnamefont {S.}~\bibnamefont {Campbell}},\ and\ \bibinfo {author}
  {\bibfnamefont {N.~T.}\ \bibnamefont {Zinner}},\ }\bibfield  {title}
  {\bibinfo {title} {Stable adiabatic quantum batteries},\ }\href
  {https://doi.org/10.1103/PhysRevE.100.032107} {\bibfield  {journal} {\bibinfo
   {journal} {Phys. Rev. E}\ }\textbf {\bibinfo {volume} {100}},\ \bibinfo
  {pages} {032107} (\bibinfo {year} {2019})}\BibitemShut {NoStop}%
\bibitem [{\citenamefont {Barfknecht}\ \emph {et~al.}(2019)\citenamefont
  {Barfknecht}, \citenamefont {Rasmussen}, \citenamefont {Foerster},\ and\
  \citenamefont {Zinner}}]{PhysRevB.99.144304}%
  \BibitemOpen
  \bibfield  {author} {\bibinfo {author} {\bibfnamefont {R.~E.}\ \bibnamefont
  {Barfknecht}}, \bibinfo {author} {\bibfnamefont {S.~E.}\ \bibnamefont
  {Rasmussen}}, \bibinfo {author} {\bibfnamefont {A.}~\bibnamefont
  {Foerster}},\ and\ \bibinfo {author} {\bibfnamefont {N.~T.}\ \bibnamefont
  {Zinner}},\ }\bibfield  {title} {\bibinfo {title} {Realizing time crystals in
  discrete quantum few-body systems},\ }\href
  {https://doi.org/10.1103/PhysRevB.99.144304} {\bibfield  {journal} {\bibinfo
  {journal} {Phys. Rev. B}\ }\textbf {\bibinfo {volume} {99}},\ \bibinfo
  {pages} {144304} (\bibinfo {year} {2019})}\BibitemShut {NoStop}%
\bibitem [{\citenamefont {Lindblad}(1976)}]{Lindblad1976}%
  \BibitemOpen
  \bibfield  {author} {\bibinfo {author} {\bibfnamefont {G.}~\bibnamefont
  {Lindblad}},\ }\bibfield  {title} {\bibinfo {title} {On the generators of
  quantum dynamical semigroups},\ }\href {https://doi.org/10.1007/BF01608499}
  {\bibfield  {journal} {\bibinfo  {journal} {Commun. Math. Phys.}\ }\textbf
  {\bibinfo {volume} {48}},\ \bibinfo {pages} {119} (\bibinfo {year}
  {1976})}\BibitemShut {NoStop}%
\bibitem [{\citenamefont {Breuer}\ and\ \citenamefont
  {Petruccione}(2002)}]{breuer2002theory}%
  \BibitemOpen
  \bibfield  {author} {\bibinfo {author} {\bibfnamefont {H.}~\bibnamefont
  {Breuer}}\ and\ \bibinfo {author} {\bibfnamefont {F.}~\bibnamefont
  {Petruccione}},\ }\href@noop {} {\emph {\bibinfo {title} {The theory of open
  quantum systems}}}\ (\bibinfo  {publisher} {Oxford University Press,
  Oxford},\ \bibinfo {year} {2002})\BibitemShut {NoStop}%
\bibitem [{\citenamefont {Poulsen}\ \emph
  {et~al.}(2022{\natexlab{c}})\citenamefont {Poulsen}, \citenamefont {Santos},\
  and\ \citenamefont {Zinner}}]{PhysRevLett.128.240401}%
  \BibitemOpen
  \bibfield  {author} {\bibinfo {author} {\bibfnamefont {K.}~\bibnamefont
  {Poulsen}}, \bibinfo {author} {\bibfnamefont {A.~C.}\ \bibnamefont
  {Santos}},\ and\ \bibinfo {author} {\bibfnamefont {N.~T.}\ \bibnamefont
  {Zinner}},\ }\bibfield  {title} {\bibinfo {title} {Quantum wheatstone
  bridge},\ }\href {https://doi.org/10.1103/PhysRevLett.128.240401} {\bibfield
  {journal} {\bibinfo  {journal} {Phys. Rev. Lett.}\ }\textbf {\bibinfo
  {volume} {128}},\ \bibinfo {pages} {240401} (\bibinfo {year}
  {2022}{\natexlab{c}})}\BibitemShut {NoStop}%
\bibitem [{\citenamefont {Kjaergaard}\ \emph {et~al.}(2020)\citenamefont
  {Kjaergaard}, \citenamefont {Schwartz}, \citenamefont {Braumüller},
  \citenamefont {Krantz}, \citenamefont {Wang}, \citenamefont {Gustavsson},\
  and\ \citenamefont {Oliver}}]{doi:10.1146/annurev-conmatphys-031119-050605}%
  \BibitemOpen
  \bibfield  {author} {\bibinfo {author} {\bibfnamefont {M.}~\bibnamefont
  {Kjaergaard}}, \bibinfo {author} {\bibfnamefont {M.~E.}\ \bibnamefont
  {Schwartz}}, \bibinfo {author} {\bibfnamefont {J.}~\bibnamefont
  {Braumüller}}, \bibinfo {author} {\bibfnamefont {P.}~\bibnamefont {Krantz}},
  \bibinfo {author} {\bibfnamefont {J.~I.-J.}\ \bibnamefont {Wang}}, \bibinfo
  {author} {\bibfnamefont {S.}~\bibnamefont {Gustavsson}},\ and\ \bibinfo
  {author} {\bibfnamefont {W.~D.}\ \bibnamefont {Oliver}},\ }\bibfield  {title}
  {\bibinfo {title} {Superconducting qubits: Current state of play},\ }\href
  {https://doi.org/10.1146/annurev-conmatphys-031119-050605} {\bibfield
  {journal} {\bibinfo  {journal} {Annu. Rev. Condens. Matter Phys.}\ }\textbf
  {\bibinfo {volume} {11}},\ \bibinfo {pages} {369} (\bibinfo {year}
  {2020})}\BibitemShut {NoStop}%
\bibitem [{\citenamefont {Lawrie}\ \emph {et~al.}(2020)\citenamefont {Lawrie},
  \citenamefont {Hendrickx}, \citenamefont {van Riggelen}, \citenamefont
  {Russ}, \citenamefont {Petit}, \citenamefont {Sammak}, \citenamefont
  {Scappucci},\ and\ \citenamefont {Veldhorst}}]{Lawrie2020}%
  \BibitemOpen
  \bibfield  {author} {\bibinfo {author} {\bibfnamefont {W.~I.~L.}\
  \bibnamefont {Lawrie}}, \bibinfo {author} {\bibfnamefont {N.~W.}\
  \bibnamefont {Hendrickx}}, \bibinfo {author} {\bibfnamefont {F.}~\bibnamefont
  {van Riggelen}}, \bibinfo {author} {\bibfnamefont {M.}~\bibnamefont {Russ}},
  \bibinfo {author} {\bibfnamefont {L.}~\bibnamefont {Petit}}, \bibinfo
  {author} {\bibfnamefont {A.}~\bibnamefont {Sammak}}, \bibinfo {author}
  {\bibfnamefont {G.}~\bibnamefont {Scappucci}},\ and\ \bibinfo {author}
  {\bibfnamefont {M.}~\bibnamefont {Veldhorst}},\ }\bibfield  {title} {\bibinfo
  {title} {Spin relaxation benchmarks and individual qubit addressability for
  holes in quantum dots},\ }\href
  {https://doi.org/10.1021/acs.nanolett.0c02589} {\bibfield  {journal}
  {\bibinfo  {journal} {Nano Lett.}\ }\textbf {\bibinfo {volume} {20}},\
  \bibinfo {pages} {7237} (\bibinfo {year} {2020})}\BibitemShut {NoStop}%
\bibitem [{\citenamefont {Häffner}\ \emph {et~al.}(2008)\citenamefont
  {Häffner}, \citenamefont {Roos},\ and\ \citenamefont
  {Blatt}}]{HAFFNER2008155}%
  \BibitemOpen
  \bibfield  {author} {\bibinfo {author} {\bibfnamefont {H.}~\bibnamefont
  {Häffner}}, \bibinfo {author} {\bibfnamefont {C.}~\bibnamefont {Roos}},\
  and\ \bibinfo {author} {\bibfnamefont {R.}~\bibnamefont {Blatt}},\ }\bibfield
   {title} {\bibinfo {title} {Quantum computing with trapped ions},\ }\href
  {https://doi.org/https://doi.org/10.1016/j.physrep.2008.09.003} {\bibfield
  {journal} {\bibinfo  {journal} {Phys. Rep.}\ }\textbf {\bibinfo {volume}
  {469}},\ \bibinfo {pages} {155} (\bibinfo {year} {2008})}\BibitemShut
  {NoStop}%
\bibitem [{\citenamefont {Saffman}\ \emph {et~al.}(2010)\citenamefont
  {Saffman}, \citenamefont {Walker},\ and\ \citenamefont
  {M\o{}lmer}}]{RevModPhys.82.2313}%
  \BibitemOpen
  \bibfield  {author} {\bibinfo {author} {\bibfnamefont {M.}~\bibnamefont
  {Saffman}}, \bibinfo {author} {\bibfnamefont {T.~G.}\ \bibnamefont
  {Walker}},\ and\ \bibinfo {author} {\bibfnamefont {K.}~\bibnamefont
  {M\o{}lmer}},\ }\bibfield  {title} {\bibinfo {title} {Quantum information
  with rydberg atoms},\ }\href {https://doi.org/10.1103/RevModPhys.82.2313}
  {\bibfield  {journal} {\bibinfo  {journal} {Rev. Mod. Phys.}\ }\textbf
  {\bibinfo {volume} {82}},\ \bibinfo {pages} {2313} (\bibinfo {year}
  {2010})}\BibitemShut {NoStop}%
\bibitem [{\citenamefont {Rasmussen}\ \emph {et~al.}(2021)\citenamefont
  {Rasmussen}, \citenamefont {Christensen}, \citenamefont {Pedersen},
  \citenamefont {Kristensen}, \citenamefont {B\ae{}kkegaard}, \citenamefont
  {Loft},\ and\ \citenamefont {Zinner}}]{PRXQuantum.2.040204}%
  \BibitemOpen
  \bibfield  {author} {\bibinfo {author} {\bibfnamefont {S.~E.}\ \bibnamefont
  {Rasmussen}}, \bibinfo {author} {\bibfnamefont {K.~S.}\ \bibnamefont
  {Christensen}}, \bibinfo {author} {\bibfnamefont {S.~P.}\ \bibnamefont
  {Pedersen}}, \bibinfo {author} {\bibfnamefont {L.~B.}\ \bibnamefont
  {Kristensen}}, \bibinfo {author} {\bibfnamefont {T.}~\bibnamefont
  {B\ae{}kkegaard}}, \bibinfo {author} {\bibfnamefont {N.~J.~S.}\ \bibnamefont
  {Loft}},\ and\ \bibinfo {author} {\bibfnamefont {N.~T.}\ \bibnamefont
  {Zinner}},\ }\bibfield  {title} {\bibinfo {title} {Superconducting circuit
  companion---an introduction with worked examples},\ }\href
  {https://doi.org/10.1103/PRXQuantum.2.040204} {\bibfield  {journal} {\bibinfo
   {journal} {PRX Quantum}\ }\textbf {\bibinfo {volume} {2}},\ \bibinfo {pages}
  {040204} (\bibinfo {year} {2021})}\BibitemShut {NoStop}%
\bibitem [{\citenamefont {Devoret}\ and\ \citenamefont
  {Schoelkopf}(2013)}]{doi:10.1126/science.1231930}%
  \BibitemOpen
  \bibfield  {author} {\bibinfo {author} {\bibfnamefont {M.~H.}\ \bibnamefont
  {Devoret}}\ and\ \bibinfo {author} {\bibfnamefont {R.~J.}\ \bibnamefont
  {Schoelkopf}},\ }\bibfield  {title} {\bibinfo {title} {Superconducting
  circuits for quantum information: An outlook},\ }\href
  {https://doi.org/10.1126/science.1231930} {\bibfield  {journal} {\bibinfo
  {journal} {Science}\ }\textbf {\bibinfo {volume} {339}},\ \bibinfo {pages}
  {1169} (\bibinfo {year} {2013})}\BibitemShut {NoStop}%
\bibitem [{\citenamefont {Ronzani}\ \emph {et~al.}(2018)\citenamefont
  {Ronzani}, \citenamefont {Karimi}, \citenamefont {Senior}, \citenamefont
  {Chang}, \citenamefont {Peltonen}, \citenamefont {Chen},\ and\ \citenamefont
  {Pekola}}]{Ronzani2018}%
  \BibitemOpen
  \bibfield  {author} {\bibinfo {author} {\bibfnamefont {A.}~\bibnamefont
  {Ronzani}}, \bibinfo {author} {\bibfnamefont {B.}~\bibnamefont {Karimi}},
  \bibinfo {author} {\bibfnamefont {J.}~\bibnamefont {Senior}}, \bibinfo
  {author} {\bibfnamefont {Y.-C.}\ \bibnamefont {Chang}}, \bibinfo {author}
  {\bibfnamefont {J.~T.}\ \bibnamefont {Peltonen}}, \bibinfo {author}
  {\bibfnamefont {C.}~\bibnamefont {Chen}},\ and\ \bibinfo {author}
  {\bibfnamefont {J.~P.}\ \bibnamefont {Pekola}},\ }\bibfield  {title}
  {\bibinfo {title} {Tunable photonic heat transport in a quantum heat valve},\
  }\href {https://doi.org/10.1038/s41567-018-0199-4} {\bibfield  {journal}
  {\bibinfo  {journal} {Nat. Phys.}\ }\textbf {\bibinfo {volume} {14}},\
  \bibinfo {pages} {991} (\bibinfo {year} {2018})}\BibitemShut {NoStop}%
\end{thebibliography}%

\end{document}